# TAGIFY: LLM-powered Tagging Interface for Improved Data Findability on OGD portals

Kevin Kliimask, Anastasija Nikiforova,
*Institute of Computer Science, University of Tartu*, Tartu, Estonia
Email: kevin.kliimask@ut.ee, nikiforova.anastasija@gmail.com

**Abstract**: Efforts directed towards promoting Open Government Data (OGD) have gained significant traction across various governmental tiers since the mid-2000s. As more datasets are published on OGD portals, finding specific data becomes harder, leading to information overload and so-called "dark data". Complete and accurate documentation of datasets, including association of proper tags with datasets is key to improving dataset findability and accessibility. Analysis conducted on the Estonian Open Data Portal revealed that 11% datasets have no associated tags, while 26% had only one tag assigned to them, which underscores challenges in data findability and accessibility within the portal, which, according to the recent Open Data Maturity Report, is considered trend-setter. The aim of this study is to propose an automated solution to tagging datasets to improve data findability on OGD portals. This paper presents TAGIFY – a prototype of tagging interface that employs large language models (LLM) such as GPT-3.5-turbo and GPT-4 to automate dataset tagging, generating tags for datasets in English and Estonian, thereby augmenting metadata preparation by data publishers and improving data findability on OGD portals by data users. The developed solution was evaluated by users and their feedback was collected to define an agenda for future prototype improvements.

## 1. Introduction

In recent times, governments around the world have developed Open Government Data (OGD) initiatives, merging the principles of open government and open data [1]. Governments around the globe publish governmental data on OGD portals that become available for a wide audience of users. Yet, users face difficulties in finding and discovering datasets related to their goals due to lack of sufficient descriptive metadata in open data catalogues [2]. As an increasing number of datasets become accessible, the challenge of information overload arises due to the difficulty in locating specific information [3], often resulting in "dark data". A recommended approach to enhance the discoverability and shareability of published datasets involves employing expressive descriptors, such as tags, effectively [3]. Descriptors constitute a form of metadata, providing details about the content of a resource to assist in its discovery or comprehension [4]. Incomplete or inaccurate metadata inhibits consumers from discovering relevant data for their requirements, leading to the necessity of spending significant time manually searching through portals and the data itself to identify relevant datasets [3, 5].

While tags may seem to be a trivial searching facet, the current practice shows that both their presence and relevance to the actual dataset tend to be a challenge for OGD portals, including the Estonian Open Data Portal, which, according to various open data rankings, is recognized among "trend-setters" [20]. Through an analysis of the Estonian Open Data Portal conducted as part of this study (Section II.C), significant shortcomings were discovered in data tagging practices, i.e., among 1787 datasets published (as of April 2024), 11% have no associated tags, while for 26% datasets have only one tag assigned to them. These findings underscore challenges in data findability and accessibility within the portal.

The aim of this study is to address the challenges associated with dataset findability and metadata quality by developing a prototype that automates the tagging process by employing a large language model (LLM). The emergence of LLMs as a subset of Generative AI offers a compelling alternative, utilizing advanced natural language understanding to generate meaningful tags. The development of such a tool holds significant promise for both data publishers and consumers. Automating the tagging process for data publishers mitigates the risk of datasets lacking tags, a common occurrence on portals where their inclusion isn't mandatory, i.e., the design of the portal does not suggest that they are mandatory. Moreover, this automation minimises the association of datasets with incomplete or inaccurate tags. Consequently, it enhances the findability and accessibility of datasets, facilitating streamlined access for users. By streamlining the metadata enrichment process, publishers can allocate resources more efficiently and accelerate the dissemination of datasets associated with high-quality tags.

As such, this study proposes TAGIFY – LLM-powered TAGging Interface to automate the tagging process for datasets formatted in CSV as one of the most popular open data formats [6, 7]. TAGIFY is developed as a web service, designed to optimise interoperability and integration with different platforms and systems. Additionally, a front-end application is developed to conduct its usability testing with actual users. The prototype is evaluated with over 20 users, collecting the feedback on its efficiency, effectiveness, evaluating the relevancy of generated tags, along with user-friendliness and usefulness of the prototype, as well as collecting the feedback for its further improvement.

The paper is organised as follows: Section 2 defines the core concepts related to the study and explains data findability issues faced by users of open data portals. Section 3 introduces the implementations strategy of this study. Section 4 presents the implementation of the prototype. Section 5 presents evaluation of the prototype. Section 6 and 7 presents the discussion, acknowledge limitations, and outlines future improvements of the prototype.

## 2. Background

This section defines the core concepts used in the paper, including OGD and FAIR principles with further determination of the problem this study attempts to resolve.

### 2.1. Open Government Data vs FAIR data

Initiatives aimed at fostering OGD, including the establishment of OGD portals, have seen widespread adoption since the mid-2000s across governmental levels [8]. The Organisation for Economic Co-operation and Development (OECD) defines OGD as both a philosophy and a set of policies aimed at fostering transparency, accountability, and value generation by making government data accessible to the public [9]. To be recognized as OGD these data must be also compliant with principles set by the Open Data Charter, according to which they must be open by default, timely and comprehensive, accessible and usable, comparable and interoperable, suitable for improved governance and citizen engagement, as well as for inclusive development and innovation [10]. By sharing data that public entities generate in substantial amounts, these data are seen to enhance transparency and accountability to citizens, while their use and reuse promote the creation of businesses and innovative citizen-centric services [9]. This movement has been joined by most countries globally.

Another concept closely related to OGD that aims to maximise the value and usability of data, albeit within another context, is FAIR. FAIR [12] is the set of guiding principles that enable both machines and humans to find, access, interoperate and re-use data and metadata [13,14]. FAIR stands for Findability, Accessibility, Interoperability and Reusability. Findability is the principle, according to which, both humans and computers should encounter minimal difficulty in locating metadata and data resources, where machine-readable metadata plays a crucial role in facilitating automated discovery of datasets and services, thus constituting a fundamental aspect of the FAIRification process. Accessibility requires that after locating the desired data, the user must ascertain the methods for accessing them, which may involve considerations such as authentication and authorization processes. Interoperability sets prerequisites for data to be integrated with other datasets, making them capable of interoperating with various applications or workflows for purposes such as analysis, storage, and processing. Reusability, being the primary goal of FAIR, dictates the need to enhance the efficiency of data reuse. This entails ensuring that metadata and data are well-described so they can be reused in different settings [14].

As such, OGD initiatives and FAIR principles share common goals of maximising the value and usability of data by promoting principles of openness, accessibility, interoperability, and reusability. However, although both are related concepts, they serve slightly different purposes, where OGD initiatives focus specifically on making government data open and accessible to the public, while FAIR principles provide a broader framework for ensuring that data, regardless of its source, is findable, accessible, interoperable, and reusable (FAIR). As such, data can be compliant with the open (government) data principles, but not necessarily compliant with FAIR principles and vice versa, FAIR data is not necessarily open (government) data principles-compliant, whereas the greatest result is achieved, when both sets of principles are fulfilled [15,42].

### 2.2. Data Findability Issues

Data published on (open) data portals is subject for search through several approaches, namely, text search that allows dataset search by their title, and faceted search that allows datasets search by facets such as publisher, file format, spatial/geographical coverage, time period-, keyword- and tag-based, with keyword- and tag-based search being prevalent [1,16]. While some facets used for dataset search can be automatically retrieved from the data associated with the publisher or the dataset, e.g., dataset format, some facets, such as spatial coverage and tags, are expected to be provided by the data publisher, where the quality of tags (completeness, accuracy etc.), depend directly on the data provided by publishers. These tags being thought of as "expressive descriptors" [3] play a crucial role in facilitating efficient navigation through data portals [2, 3]. By associating datasets with relevant tags, users can locate datasets relating to specific topics of their interest [2, 3].

Entering tags manually is slow and prone to human errors, such as tags are not always being accurate or relevant to the actual dataset [3, 17]. Additionally, if the portal's design doesn't enforce mandatory tagging, publishers may overlook tagging entirely due to its time-consuming nature, which is one of barriers towards data opening [20]. Inadequate metadata, including descriptions or tags, renders both manual and automated searches ineffective in locating the dataset, thus making the dataset non-findable, inaccessible, interoperable, and consequently – non-re-usable [18,42].

While tags may seem to be a trivial facet, the current practice shows that both their presence and relevance to the actual dataset tend to be a challenge for OGD portals, including the Estonian Open Data Portal, as found out in conversation with Estonian Open Data Portal representatives, regardless of the fact that, according to various open data rankings, is recognized among "trend-setters" [20]. To this end, to assess the relevance of the topic of this study and Estonian open data as a domain of application, an analysis of datasets available on the Estonian Open Data Portal was conducted with the aim to examine the relevance of the issue in question, i.e., lack of or insufficient quality of tags associated with published datasets on Estonian Open Data Portal.

### 2.3. Analysis of datasets tags in Estonian Open Data Portal

To analyse the number of tags associated with each dataset on the portal as defined by data publishers, a Python scraping script was developed (see Fig. 1), the code of which is available in a Github repository[1]. The script operates as follows:

- list of all datasets on the portal is retrieved using the *get_datasets_list()* function, which iterates over each dataset. *get_datasets_list()* fetches datasets from the API endpoint https://avaandmed.eesti.ee/api/datasets;

---

[1] https://github.com/kevinkliimask/gpt-tagger

- for each dataset, information is retrieved using the *get_dataset(uuid)* function, where the parameter uuid is the dataset's unique identifier. It fetches the detailed information from the API endpoint *https://avaandmed.eesti.ee/api/datasets/{uuid}*;
- the length of the datasets's tag/keywords field is determined. The count of tags for each dataset is then incremented in the counts dictionary;
- once all the retrieved datasets are processed, the dictionary containing counts for each number of tags is printed out.

```python
import requests
import json
from collections import import defaultdict

counts = defaultdict(lambda : 0)

def get_datasets_list():
    x = requests.get('https://avaandmed.eesti.ee/api/datasets?limit=1787')
    return json.loads(x.content)['data']

def get_dataset(uuid):
    x = requests.get(f'https://avaandmed.eesti.ee/api/datasets/{uuid}')
    return json.loads(x.content)['data']

if __name__ == '__main__':
    data = get_datasets_list()
    i = 0

    try:
        for obj in data:
            dataset = get_dataset(obj['id'])
            counts[len(dataset['keywords'])] += 1
            i += 1
    except:
        print(counts)

    print(counts)
```

Fig. 1. Estonian open data portal scraping script

The analysis performed in accordance with this procedure, uncovered significant negative trends in datasets tagging practice, thereby confirming the relevance of the study objective. Out of the 1787 datasets published (as of 23.04.24), 190 datasets (11%) lacked any associated tags, while 457 (26%) had only one tag assigned to them. This infringes the principles of FAIR, i.e., if a dataset is lacking relevant metadata such as tags, it will be more difficult for interested parties to find it (infringes findability) and to integrate it with other datasets (infringes interoperability) [14]. As a result, lack of keywords/tags will make the dataset less likely to be reused, which infringes reusability [14]. This indicates potential areas for improvement in dataset tagging within the portal through augmentation of this process, which is a central objective of this study. As advancements in artificial intelligence technologies continue, they can be harnessed to enhance the discoverability of data through automated tagging of datasets. Moreover, automation elements are inherent to the FAIR vision [19].

### 3. Implementation Strategy and Technological Framework

The objective of the study is achieved by automating dataset tagging, which, in turn, is achieved by employing a LLM that powers TAGIFY – a prototype of tagging interface. LLM is appropriate for this purpose, as it was found to be useful for predicting tags from partial content of a dataset [22]. As such, the following steps outline the process of automatically tagging a dataset:

1. the LLM gets a system prompt describing to it which data it will receive, which task it has to do and how its response should be formatted. A system prompt is a message that can be used to specify the persona used by the model in its replies [23]. Instructions to the LLM are provided in English;
2. then, the LLM is provided with the first rows of a dataset, including the dataset's header row. The number of rows provided to the LLM is 10. Experimentation has shown that this number of rows is one of the lowest that still allows the LLM to generate relevant tags. Moreover, every additional row provided for analysis would increase the computational resources required, thus making the process more expensive. Furthermore, this number of rows also fits inside the input token limit of the LLM, which determines the maximum length of the input string that the LLM can accept;
3. after processing the input, the LLM outputs a list of relevant tags. The tags are in English. The number of tags to output can be chosen by the user, of which the LLM is informed through the initial system prompt;
4. finally, in addition to the English tags generated by LLM at step 3, translation of the generated tags in Estonian, as the language of a portal with which it will be tested and to which it is planned to integrate it to, is returned to the user. In other words, translations do not originate from the LLM, instead, the English tags generated by the LLM are translated separately by using a machine translation service's API.

### 3.1. Interfacing with the LLM

Communication with the LLM is achieved through a RESTful web service, which handles interfacing with the LLM's API. A RESTful web service is a web application that adheres to REST standards [24]. Web services enable various organisations or applications from diverse origins to interact without the necessity of exchanging sensitive data or IT infrastructure [25]. Developing the project as a web service has the benefit of not limiting the project to the Estonian Open Data Portal or OGD portals in general, thereby making it environment-agnostic, which will make it convenient to integrate the tagging service with other products.

Additionally, a basic graphical user interface (GUI) is developed to interface with the web service. This is done to allow for a more streamlined and user-friendly usability testing (section 5). In order to facilitate usability testing over distance, the application is deployed to the cloud. By implementing this approach, users will be spared the need to set up the application locally, thus alleviating the associated inconvenience. Furthermore, it ensures that sensitive API keys remain protected and do not need to be shared with users during the testing phase.

### 3.2. Technology Choices

This section presents technological choices made to develop an automated tagging service, with the reference to both LLM, web service framework, GUI framework, translation service and cloud provider.

*1) **Large Language Model***: Since the prototype under development is LLM-powered, the first technological choice concerned which LLM to use. The factors that determined the choice of LLM were performance, cost and ease of implementation. Several benchmarks have been developed to evaluate the performance of a LLM, such as *HELM (Holistic Evaluation of Language Models)*, which is a research benchmark developed by the Stanford CRFM (Center for Research on Foundation Models) to assess performance across a variety of prediction and generation scenarios, *Open LLM leaderboard* by HuggingFace, which is a leaderboard for open source LLM evaluation across 4 benchmarks - *MMLU*, *TruthfulQA*, *HellaSwag* and *AI2* reasoning, and *Chatbot Arena* by LMSys, which is a benchmark utilising an Elo-derived ranking system, aggregated over pairwise battles [26]. However, there is no widely used benchmark for evaluating performance of LLMs as data annotators [26]. To this end, we referred to a technical report by Refuel [26] to find a LLM with the best trade-off between label quality and cost. The report evaluated the performance of 6 LLMs, namely Text-davinci-003, GPT-3.5-turbo, GPT-4, Claude-v1, FLAN-T5-XXL, PaLM-2 for labelling datasets. The report identified that the top-3 LLMs with the best trade-off between label quality and cost are FLAN-T5-XXL, PaLM-2 and GPT-3.5-turbo, which were further considered for the purpose of this study. An additional investigation of the three LLMs revealed that FLAN-T5-XXL requires self-hosting, which increases the complexity of developing the solution. PaLM-2 and GPT-3.5-turbo offer a paid API, which is easier to implement than a self-hosted LLM. As the performance of the 2 models is similar, where GPT-3.5-turbo generates better quality labels compared to PaLM-2 in 5 out of 10 datasets, while the cost per label of PaLM-2 is ~70% higher [21], the choice to use GPT-3.5-turbo was made.

In addition, during development, the decision to include GPT-4 as an additional option was made, incl. to better evaluate GPT-3.5-turbo.

*2)* **Web Service**: A RESTful web framework was used to develop a HTTP-based API for accessing the web service. The choice of framework was FastAPI - a *"web framework for building APIs with Python 3.8+ based on standard Python type hints"* [27], as according to independent benchmarks by TechEmpower, FastAPI is considered as one of the fastest Python frameworks available, only below Starlette and Uvicorn [28]. FastAPI is built upon Starlette, which itself is built upon Uvicorn, which explains the differences in performance as this hierarchical architecture inherently introduces additional layers of abstraction, resulting in increased overhead [28]. But as an added benefit, FastAPI provides more features on top of Starlette, such as data validation and serialisation that are essential to building APIs [28]. By using a higher-level framework such as FastAPI, development time is saved and similar performance to a lower-level framework, such as Starlette, can be achieved as features missing in Starlette would have to be developed manually [28]. In addition, OpenAI (the company that offers the GPT-3.5-turbo and GPT-4 models) provides official Python bindings for using their models [29], which makes using a Python-based framework convenient.

*3)* **Graphical User Interface**: When making a decision about a graphical user interface, a choice in favour of one of two options should be made, namely a desktop application or a web application. A front-end web application as the graphical user interface was chosen to facilitate a more seamless user-testing experience. The decision was influenced by several factors. Notably, web applications offer the advantage of immediate accessibility without the need for installation, ensuring users can swiftly engage with the application across different devices and operating systems [30]. While it is acknowledged that web applications rely on an internet connection, which could be perceived as a limitation [30], usage of the LLM requires an Internet connection regardless. Therefore, this potential drawback becomes irrelevant in the context of this study.

*Node.js* and *React* stand as two of the most used front-end web frameworks globally [31]. Node.js is an open-source JavaScript runtime environment that facilitates the development of servers and web applications [32]. Conversely, React is described as a *"library for web and native user interfaces"* [33]. Given that Node.js is predominantly tailored towards API creation, while React is renowned for its prowess in creating user interfaces [34], the decision to use React was made due to its better alignment with the project's requirements.

*4)* **Machine Translation Service**: For this project, the criteria for choosing a machine translation service was that it must be accurate and have an accessible API. According to research conducted by Intento [35], DeepL emerged as the top-performing neural machine translation service. DeepL offers a free, although limited, access plan to access their API. Additionally, the existence of an official Python library maintained by DeepL facilitates its convenient integration into the application. Considering these factors, the decision to use DeepL as the project's machine translation service was made. Although Google Translate was initially considered during the project's early stages, a comparative analysis revealed that DeepL consistently delivered more accurate translations. This performance disparity ultimately solidified DeepL as the preferred choice for the project.

*5)* **Cloud Provider**: When selecting a cloud provider, the primary criteria were cost-effectiveness and ease of application deployment. For this project Vercel was chosen. Vercel is a cloud-based platform specifically tailored for hosting static sites and serverless functions, offering developers a streamlined process in developing and launching web projects [37]. Vercel offers the ability to run back-end code as serverless functions [37]. A serverless function embodies business logic that operates without retaining data (stateless) and has a temporary lifespan, being created and then terminated [38]. These functions persist for short durations, mere seconds, and are intended to be triggered by a specific condition, such as an user making a request. Given that the web service does not need to retain data and only needs to run upon a request, the utilisation of serverless functions was deemed aligned with the project. In addition, Vercel offers a free tier and its straightforward deployment process further solidified its suitability.

## 4. Implementation

In this section, implementation of the prototype back-end is presented, with subsequent presentation of the front-end. Finally, the process of hosting the application is presented.

### 4.1. Back-end

The back-end of the developed prototype consists of 3 main modules: (1) API endpoint, (2) OpenAI service and (3) translator service, each playing its own crucial role:

- **API endpoint** accepts requests and validates received data from the user, which is expected to be a .csv file. The data consists of the first 10 rows of a *to-be tagged* dataset, including its header row;
- **OpenAI** service handles interfacing with OpenAI API. It creates a system prompt, appends data received from the API endpoint to a user prompt and sends both messages to the LLM. It, in turn, receives a response from the LLM with tags (generated by the LLM in English);
- **Translator** service handles interfacing with DeepL API. It takes tags received from OpenAI service and translates them to another language, which within the context of the study is Estonian;
- **Config** module handles loading environment variables into the application. The necessary environment variables for the back-end application to function are front-end url, OpenAI API key and DeepL API key.

In addition, the back-end project contains a requirements file for required Python packages and a Vercel configuration file for the prototype application deployment purposes. The required Python packages for the project are *fastapi*, *pydantic*, *pydantic-settings*, *python-multipart*, *uvicorn*, *openai*, *deepl* and all dependencies of the preceding packages. All parts of the developed prototype are available in a Github repository[2]. In subsequent subsections each module is presented in more detail.

*1)* **API Endpoint**: The API endpoint accepts data sent via HTTP POST method. Furthermore, the endpoint is mapped to the "/" route, also known as the root route. As there are no other endpoints in the application, it is sufficient to accept requests only on the root route.

Received data is validated to prevent unexpected behaviours in the application. The API endpoint accepts a body consisting of a matrix, where the matrix represents data from a dataset. In addition, the endpoint accepts count and model as query parameters from the user. These determine how many tags the LLM should generate and which LLM model should be used, respectively. The default values for these parameters are 5 tags and GPT-3.5-turbo model. The validation logic sets the following rules for the received data:

- length of data in the request body, which represents the number of rows of a dataset, must be a maximum of 10 lines;

---

[2] https://github.com/kevinkliimask/gpt-tagger

- count should be in the range of 3 to 10;
- model should be either GPT-3.5-turbo or GPT-4.

If any of the validations fail, a HTTP exception is returned as a response shown to the user, specifying the nature of the error (see Fig. 2).

```python
class Data(BaseModel):
    data: List[List[str]]

@app.post("/")
async def get_tags(data: Data, count: int = 5, model: str = "gpt-3.5-turbo"):
    if len(data.data) > 10:
        raise HTTPException(status_code=400, detail="Data length must be a maximum of 10 lines")
    if not 3 <= count <= 10:
        raise HTTPException(status_code=400, detail="Count must be between 3 and 10")
    if not model in ["gpt-3.5-turbo", "gpt-4"]:
        raise HTTPException(status_code=400, detail="Model must be gpt-3.5-turbo or gpt-4")

    return await handle_tagging(data.data, count, model)
```

Fig. 2. API endpoint and data validation logic.

*2) OpenAI Service*: The OpenAI service defines a function *handle_tagging* that uses OpenAI API to generate tags for a dataset. Communication with OpenAI API is handled by OpenAI Python library. The function takes a list of records from a dataset, the number of tags to generate, and the model to use as input parameters, all provided by the user. The function builds messages to send to the OpenAI API, formats the data into a user message, sends the messages to the API, retrieves the generated tags, splits them into English tags, and then translates them into Estonian using translator service. Finally, it returns a dictionary containing both English and Estonian tags (see Fig. 3).

```python
client = OpenAI(api_key=settings.chatgpt_api_key)

class TagsData(TypedDict):
    english: List[str]
    estonian: List[str]

async def handle_tagging(data: List[str], count: int, model: str) -> Dict[Literal["data"], TagsData]:
    messages = [{"role": "system", "content": "You will generate tags for a dataset. I will provide your the first rows "
                                               "of the dataset, whereas the very first row will be the column titles of the dataset. "
                                               "The first row will be in the following form: title1,title2,title3,etc... "
                                               "The next rows will be in the following form: value1,value2,value3,etc... "
                                               f"Output {count} tags that describe the dataset best. Output only the "
                                               "suitable tags in the form of: tag1,tag2,tag3,etc... Tags should be in English. "
                                               f"Try to make the tags general but relevant. Output only {count} tags."}]

    user_message = ""
    for row in data:
        user_message += (",".join(row) + "\n")
    messages.append({"role": "user", "content": user_message})

    response = client.chat.completions.create(
        model=model,
        messages=messages)
    english_tags = re.split(r"\s?,\s?", response.choices[0].message.content)
    estonian_tags = translator_service.translate_text(english_tags, src="en", dest="et")

    return {"data": {"english": english_tags, "estonian": estonian_tags}}
```

Fig. 3. OpenAI service logic.

*3) Translator Service:* To ensure tags are generated in a language other than English, such as Estonian, as is the case for this study, translator service is used. The service defines a function *translate_text* that uses DeepL API to translate tags originally generated by the LLM. Interfacing with DeepL API is handled by the DeepL Python package. The function accepts a list of tags, source language and destination language as input parameters, which in this case are English and Estonian, respectively. The function translates every string in the input list and returns the translated strings as a list (see Fig. 4).

```
 8      translator = deepl.Translator(settings.deepl_auth_key)
 9
10      def translate_text(text: List[str], src: str, dest: str) -> List[str]:
11          results = translator.translate_text(text, source_lang=src, target_lang=dest)
12          return [result.text for result in results]
```

Fig. 4. Translator service logic.

*4) Config module:* The config module defines a Settings class that inherits from *BaseSettings* provided by Pydantic - a library for data validation and settings management. It specifies the environment variables required for the application, namely *frontend_url*, *chatgpt_api_key*, and *deepl_auth_key*. Then, it creates an instance of the *Settings* class to load the values of these environment variables (see Fig. 5). This approach ensures that the application's settings are correctly loaded and validated from the environment. Additionally, this setup enables anybody to run the application and use their own environment variables seamlessly.

```
 1      from pydantic_settings import BaseSettings
 2
 3
 4      class Settings(BaseSettings):
 5          frontend_url: str
 6          chatgpt_api_key: str
 7          deepl_auth_key: str
 8
 9
10      settings = Settings()
```

Fig. 5. Config module.

### 4.2. Front-end

The front-end architecture is centred around a single React component named App, functioning as the primary entry point for the application.

*1) Dependencies:* The application's required packages are defined in a *package.json* file. For a React app to operate, the main dependencies are *react*, *react-dom* and *react-scripts*. In addition, the developed React application also makes use of the react-drag-drop-files package to handle file uploads through drag-and-drop functionality. When starting the app, the environment variable *REACT_APP_BACKEND_URL* must be defined to specify the backend server's URL.

*2) State Management:* The *useState* hook from React is used to manage component state. The App component utilises the *useState* hook to manage states of tags, *selectedNumberOfTags*, *selectedModel*, *error*, and *isLoading* (see Fig. 6). These states are essential for tracking the uploaded file, selected parameters, error messages, and loading status.

```
12  function App() {
13      const [tags, setTags] = useState(null);
14      const [selectedNumberOfTags, setSelectedNumberOfTags] = useState(NUMBER_OF_TAGS[0]);
15      const [selectedModel, setSelectedModel] = useState(MODELS[0]);
16
17      const [error, setError] = useState(null);
18      const [isLoading, setIsLoading] = useState(false);
```

Fig. 6. App state variables.

*3) **File Upload and Tag Generation***: The *handleChange* function is invoked upon uploading a file. It utilises the *readCsv* utility function to extract data from the uploaded file. The *readCsv* utility function parses the CSV file uploaded by the user, preparing the data for transmission to the backend. This function accepts a single parameter file, representing the uploaded CSV file, and returns a *Promise* resolving to an array containing the first 10 rows of the parsed CSV file data (see Fig. 7). Parsing CSV data in the front-end offers the advantage of bypassing the need to transfer large files to the back-end for processing. Consequently, this approach eliminates the need for a size limit on file uploads, with the maximum size being solely dictated by the browser, e.g., 4GB limit in Chrome.

```
1   // This function was written with the help of ChatGPT https://chat.openai.com/share/5512d600-82ba-4343-97bf-1dd6bcb5df7b
2   export const readCsv = (file) => {
3       return new Promise((resolve, reject) => {
4           const reader = new FileReader();
5
6           reader.onload = (event) => {
7               const content = event.target.result;
8               const lines = content.split("\n").slice(0, 10);
9               const dataArray = lines.map((line) => line.split(","));
10
11              resolve(dataArray);
12          };
13
14          reader.onerror = (error) => {
15              reject(error);
16          };
17
18          reader.readAsText(file);
19      });
20  };
```

Fig. 7. readCsv utility function.

Upon successful CSV file reading, the *postFile* function sends the parsed data along with selected parameters (*selectedNumberOfTags* and *selectedModel*) to the server for tag generation. The *postFile* function handles the transmission of data to the backend server for tag generation. It accepts the following parameters:

- data, which represents the first 10 rows of the uploaded CSV;
- *numberOfTags*, which is the number of tags the user has chosen to be generated by the LLM;
- model, which is the LLM that the user has chosen to be used for tag generation.

The *postFile* function constructs the backend URL using the provided environment variable *REACT_APP_BACKEND_URL*, appending query parameters for count (number of tags) and model. It then performs a POST request to the constructed URL using the JavaScript fetch API, which returns a *Promise*. Finally, the *Promise* is resolved and generated tags are extracted from the JSON response (see Fig. 8). The generated tags are then stored in the component state (tags), and any errors during the process are captured and displayed.

```javascript
export async function postFile(data, numberOfTags, model) {
    const url = new URL(process.env.REACT_APP_BACKEND_URL);
    url.searchParams.set("count", numberOfTags);
    url.searchParams.set("model", model);

    const response = await fetch(url, {
      method: "POST",
      headers: {
        Accept: "application/json",
        "Content-Type": "application/json",
      },
      body: JSON.stringify({ data }),
    });
    return (await response.json()).data;
}
```

Fig. 8. postFile function.

*4) User Interface*: The handleChange function is invoked upon uploading a file. It utilises the *readCsv* utility function to extract data from the uploaded file. The *readCsv* utility function parses the CSV file uploaded by the user, preparing the data for transmission to the backend. This function accepts a single parameter file, representing the uploaded CSV file, and returns a *Promise* resolving to an array containing the first 10 rows of the parsed CSV file data.

The user interface consists of a card layout containing the application title, parameter selection dropdowns, file uploader, and sections for displaying generated tags, loading status, and error messages. Dropdown menus are provided for selecting the number of tags and the model to be used for tag generation. The react-drag-drop-files library provides a *FileUploader* component that enables users to upload CSV files, restricting them to only CSV file types (see Fig. 9).

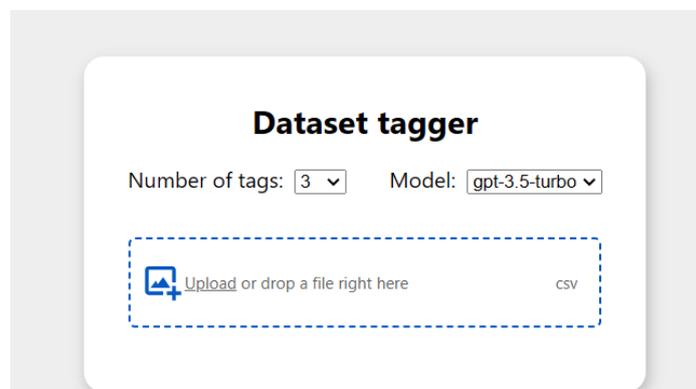

Fig. 9. Front-end application interface.

The component dynamically renders elements based on the current state. For example, it displays generated tags if available, shows loading indicators during file processing, and renders error messages if any errors occur (see Fig. 10).

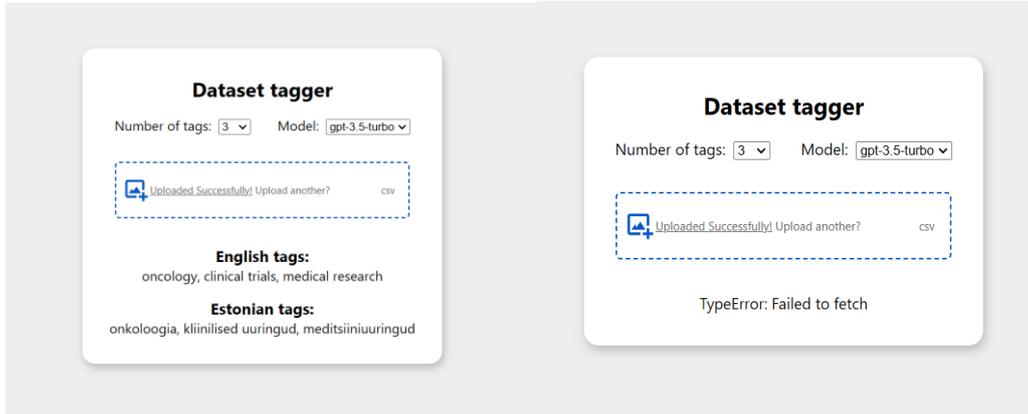

Fig. 10. Successful file upload and unsuccessful file upload.

The application's styling is maintained through CSS with style rules defined in the App.css file. These styles are designed to offer users a clean and intuitive layout, enhancing the overall interaction experience.

### 4.3. Hosting

Vercel facilitates automatic deployments triggered by changes to the respective front-end or back-end folders within the main branch of the source code's repository. Both the front-end and back-end source code are hosted in a single repository. A repository that contains multiple projects, such as the back-end and front-end, is called a *monorepo* [39]. The deployment of the application on Vercel is separated into 2 different Vercel projects: *gpt-tagger* and *gpt-tagger-frontend*. This is Vercel's recommended approach to deploying applications that use a monorepo [40].

In the case of the back-end project (*gpt-tagger*), a custom vercel.json configuration file is used to define information for Vercel to set up a Python runtime when deploying. For the front-end project (*gpt-tagger-frontend*), no configuration file is needed as Vercel can natively handle the configuration for a React application.

Vercel allows for the definition of environment variables specific to each project. As such, all necessary environment variables are defined for both projects inside the Vercel platform. The hosted prototype can be accessed via https://gpt-tagger-frontend.vercel.app/.

## 5. Evaluation of the Prototype

The primary objectives of this testing were to assess the application's functionality, relevancy of generated tags, the quality of translations from English to Estonian, user-friendliness, and gather general feedback for further improvement of the prototype. In the following subsections, the methodology and results of the evaluation are presented.

### 5.1. Prototype Evaluation Methodology

The evaluation of the prototype application involved conducting usability testing through a Google Forms survey. The survey was designed with three sections, each aimed at assessing specific aspects of the prototype, which are described below. Before taking the survey, participants were introduced with a brief description of the survey purpose (incl., its objective, brief overview of the process, and the length) and prototype, informed about consent for further use of collected data, as well as specifying that datasets uploaded are processed according to OpenAI's enterprise privacy.

The first part of the survey aimed to evaluate tagging accuracy of the prototype with pre-defined sample datasets. Participants were provided with instructions on the prototype use and links to two sample datasets sourced from the Estonian Open Data Portal. Participants were asked to try out the prototype by following the instructions on its use provided within the survey, by uploading respective datasets to the prototype. In the survey, respondents were required to answer the same set of questions for each dataset provided.

The first question was "*How relevant are the generated tags to the actual content of the datasets?*". Participants were asked to assess the relevance of the generated tags to the actual content, constituting an acceptability task, where answers were defined using 5-point Likert scale, where 1 point corresponds to "*not relevant at all*" and 5 to "*very relevant*". If a low score was assigned, the participant was followed up with an additional question asking for a justification for this score. Then, evaluation of *how the parameter "number of keywords" affects the relevancy of generated tags*. The answers were four predefined options, namely "*Yes, improves relevancy significantly*", "*Yes, improves relevancy slightly*", "*No, does not improve or worsen relevancy*" and "*No, rather worsens relevancy*". If a negative answer was given, the respondent was followed up with the open-ended question "*If tags relevancy worsens, how and at which number of keywords?*". Afterwards, the participants were asked *which LLM produced more relevant better tags* with options being "*GPT-3.5-turbo*", "*GPT-4*" and "*Both had results of similar relevancy*". Finally, the respondents were asked to assess the combination of different parameters, with the question being "*Which combination of the options "number of keywords" and "model" seemed to produce the most relevant results?*". This question was open-ended. Additionally, Estonian speakers were asked to assess the accuracy of Estonian translations of tags. As being a native speaker of Estonian was not a mandatory prerequisite for participating in the survey, this question was optional.

The second section of the survey provided participants with the opportunity to try the prototype application with their own datasets. While this section was optional, participants were encouraged to test the application with a dataset of their own choice, while providing links to Estonian Open Data Portal and European Data Portal, from which open dataset could be selected by them. After testing their dataset, participants were asked to share any observations or feedback they have regarding the tagging process. This feedback was collected to map potential areas of the prototype for improvement.

The third section of the survey focused on gathering general feedback on the prototype application. Participants were asked to provide feedback on the overall user-friendliness of the application, whether they would consider using it in their workflow (developing questions following the Unified Theory of Acceptance and Use of Technology (UTAUT) and Technology Acceptance Model (TAM) constructs for evaluating technology adoption), and if they encountered any prototype operation errors or issues during testing. In the first question of this section, the participant was asked to rate *how user-friendly the prototype is*, representing an acceptability task, with the answers defined using a 5-point Likert scale, where 1 point corresponds to "*Not user friendly*" and 5 points corresponds to "*Very user friendly*". If a low score was given, the respondent was followed up with an open-ended question to specify why they found the prototype to not be user friendly. Then, the participant was asked to rate *the usefulness of the prototype*, also constituting an acceptability task using a 5-point Likert scale, where 1 point corresponds to "*Not useful at all*" and 5 points corresponds to "*Very useful*". If the respondent found the prototype to be insufficiently useful, they were followed up with a question asking them to *justify their answer to the previous question*. The participant was then asked *if they would use the prototype for the purpose of tagging datasets* and *if they ran into any unexpected behavior or issues when using the prototype*, with both questions being closed-ended with consisting of predefined answers "*Yes*" or "*No*". If the respondent did run into unexpected behavior or issues, they were followed up with an open-ended question asking them to *describe the issue(s)*. Finally, participants were given the opportunity to offer *suggestions for improvement or features they would like to see implemented in future iterations of the application*.

### 5.2. Evaluation Results

The survey was distributed through social media, emailing to Estonian Open Data Portal representatives and personal channels, gathering in total 22 responses. The survey was targeted at individuals, who actively work or engage with datasets within their professional or personal domains.

Most respondents found generated tags relevant to the actual content of the datasets with an average value being 4.4 of 5 points, i.e., predominantly relevant, with no 1 or 2 points received. In cases, where respondents found tags to be less relevant (3 to 4 points), reasoning behind low relevance score was justified by respondents through the fact that while most tags were relevant to the dataset, some were overly specific, failing to encapsulate the broader essence of the datasets.

About 74% of respondents reported that changing the number of tags to be generated option improves relevancy with the largest share reporting that it improves changes slightly. From the obtained open-ended question seeking to find how and at which number of keywords tags relevancy worsens, a consensus emerged that increasing the number of tags generally enhanced accuracy or provided opportunities to discern more precise tags amid less accurate ones.

The majority of respondents (65%) highlighted that the best performing model was GPT-4, with the prevailing dominance of respondents highlighted that the combination of GPT-4 and utilizing 5 or more keywords appeared to consistently yield the most relevant outcomes for respondents.

Finally, as regards the Estonian tag translations, most answers accumulated to the values of 4 and 5 with no respondents assessing it with 1 or 2.

While participants found that GPT-4 generally outperforms GPT-3.5-turbo with generating tags, it was pointed out that in some rare cases the LLM returns incomprehensible output instead of relevant tags. Furthermore, several comments were made about the Estonian translations differing when using GPT-3.5-turbo and GPT-4, although these models were not used for translation, as a separate translation service was used to translate the English tags to Estonian (see Section 3 and 4).

The prototype was found by participants useful with 54% respondents rated the usefulness of the prototype with the score 4, and 27% participants found it to be very useful, thus giving it 5 of 5 points. The reasons for lower usefulness scores were commented by respondents to be due (1) the LLM has a hallucination problem, i.e., sometimes irrelevant tags are produced; (2) a tool is standalone, whereas it would be more useful if it was integrated into an open data portal; (3) multiple different combinations of "number of keywords" and "model" must be tried in order to find optimal tags.

The prototype was found to be generally user-friendly with 64% assessing it with 4 to 5 points. For the justification of lower user-friendliness scores, participants pointed out three concerns: (1) files had to be reuploaded any time the user wanted to change parameters such as "number of keywords" or "model"; (2) prototype was limited to only one file type, namely .csv; (3) file size limit was not specified, where as regards the latter - absence of a specified size limit, the prototype was designed to operate without imposing an arbitrary file size restriction (see Section IV.B).

As regards unexpected behavior or issues with the prototype, 3 participants encountered such, where the main issue that users encountered was the prototype generating a different number of tags than was actually selected in the "number of keywords" option.

Finally, 82% participants suggested they further use of the prototype. Some participants have also provided several suggestions for further improvement of the prototype, which are: (1) possibility to approve or disprove the tags coming from the model; (2) improvement of tagging accuracy; (3) option to export results, which will be considered in the considered as further improvements of the prototype.

## 6. Discussion

The proposed LLM-powered TAGging Interface for automating the dataset tagging process confirms the strong potential of LLMs in this domain. By leveraging advanced natural language understanding, TAGIFY is able to generate meaningful and contextually relevant tags, offering a significant improvement over traditional methods. This aligns with the findings of [41], where BRYT—a hybrid approach that integrates various Natural Language Processing (NLP) techniques, namely BERT, RAKE, YAKE, TextRank, and ChatGPT—has shown superior performance in automated metadata extraction, surpassing other approaches in terms of accurately extracting keywords, themes, categories, and dataset descriptions. However, despite the promising capabilities offered by AI, research in this area remains relatively underdeveloped.

The prototype developed within this study received positive feedback from participants in several key areas. Firstly, respondents generally rated the relevance of the generated tags highly, with an average rating of 4.4 out of 5. This indicates that the prototype effectively captured the essence of the datasets. Moreover, a significant portion of participants reported that adjusting the "number of keywords" option improved tag relevancy, suggesting flexibility in fine-tuning the tagging process. Additionally, most respondents favored the GPT-4 model for its superior performance in tag generation compared to GPT-3.5-turbo that was originally selected for its superiority over other LLMs (Section III.C). The superior performance of GPT-4 compared to other models aligns with findings from Refuel's LLM Labeling Technical Report [26], according to which, GPT-4 achieved an average label quality score of 0.884 (the percentage agreement with ground truth labels), whereas GPT-3.5 scored 0.813. Additionally, our evaluation of tag relevancy—considering both GPT-3.5 and GPT-4—yielded an average score of 4.4 out of 5, indicating consistency with Refuel's findings, though further evaluation with a larger user base and expanded scope is recommended to substantiate these results. The combination of GPT-4 with five or more keywords emerged as the most effective strategy for producing relevant tags consistently. Furthermore, Estonian speakers generally expressed satisfaction with the accuracy of the Estonian tag translations. These results are aligned with [26], i.e., TAGIFY being based on LLMs, can achieve dataset labeling quality comparable to or exceeding that of skilled human annotators, while data publishers are often not skilled annotators, but signifitcantly faster and cheaper (e.g., according to [26], ~20x faster and ~7x cheaper).

Despite the positive reception, user feedback identified areas for improvement in the prototype. Notably, some participants encountered instances, where the application produced irrelevant tags or incomprehensible output. In addition, some feedback highlighted that certain tags were overly specific, failing to encapsulate the broader content adequately. Furthermore, feedback regarding user interface and functionality emphasized concerns, such as the need to reupload files when adjusting parameters, limitations in supported file types and the standalone nature of the tool.

To improve the usefulness of the application, issues with tagging accuracy (although pointed to by a minority of participants) must be addressed. These issues could be addressed by refining the initial system prompt provided to the LLM or by supplying more than 10 rows of dataset content for the LLM to analyze. Although experimentation has shown that 10 rows is one of the lowest thresholds that still allows the LLM to generate relevant tags, where every additional row provided for analysis would increase the computational resources required, thus making the process more expensive (Section III), increasing the amount of data the LLM processes allows it to make better generalizations based on the dataset.

Furthermore, feedback regarding user interface and functionality emphasised concerns such as the need to reupload files when adjusting parameters, limitations in supported file types and the standalone nature of the tool. These recommended improvements to the user interface can be implemented in the future to enhance user experience, particularly focusing on the interaction with the file upload logic. This includes expanding the file type support to common formats such as JSON, HTML, XLS, XLSX, and XML, and ensuring parameter values can be changed dynamically without the need to re-upload the dataset. The latter, namely, stand-alone nature of the tool stressed by evaluators, however, is due to the fact that the evaluated artefact is a prototype, which was made publicly available by hosting it as a stand-alone tool exclusively for its testing purposes. As such, once it is improved to meet evaluators expectations, it is expected to be integrated with existing open data portals, thereby broadening accessibility and utility for a wider audience.

## 7. Conclusion

This study aimed to address the challenge of poor data findability and metadata quality associated with open datasets. To this end, TAGIFY – LLM-powered TAGging Interface -was developed that automates the tagging process by employing GPT-3.5-turbo and GPT-4, which presents significant benefits for both data publishers and consumers. Automatic tagging can reduce the risk for data publishers of publishing datasets that lack tags, which is a common issue on portals where tag indication when preparing metadata accompanying dataset is not mandatory. Additionally, this automation reduces association of datasets with incomplete or inaccurate tags, thereby contributing to metadata quality, as well as dataset's FAIRness.

The application was developed as a web service. This approach was chosen to ensure that the project is not limited to the Estonian Open Data Portal or OGD portals in general, making it environment-agnostic and interoperable with other products.

In assessing the prototype, a survey was administered, garnering 22 responses. Participants assessed various aspects of the application through its thorough examination, including the relevance of generated tags, user-friendliness, and overall usefulness, which were generally positively assessed by them. The feedback provided by respondents was used to identify areas for future improvement of the prototype.

As such, this study contributes to the realm of open data by promoting greater transparency and documentation through the adoption of Generative AI, which is emerging as a key component of the Fourth Wave of Open Data [42]. This, in turn, improves data findability, accessibility, enabling interoperability and as a result – reusability, thereby contributing to development and maintenance of more resilient and sustainable public and open data ecosystem..

## Declaration of generative AI and AI-assisted technologies in the writing process.

During the preparation of this work the author(s) used ChatGPT-3.5 to improve the flow and language. After using this tool, the authors reviewed and edited the content as needed and take(s) full responsibility for the content of the published article.